\documentclass[twocolumn,prl,tightenlines,superscriptaddress,showpacs]{revtex4}

\usepackage{amsmath}
\usepackage{amssymb,amsfonts,latexsym}
\usepackage{bm}
\usepackage[mathcal]{euscript}
\usepackage{graphicx}
\usepackage{epsfig}
\usepackage{color}
\usepackage{xfrac}

\begin{document}

\title{Large-scale chaos and fluctuations in active nematics}

\author{Sandrine Ngo}
\affiliation{Service de Physique de l'Etat Condens\'e, CNRS URA 2464, CEA-Saclay, 91191 Gif-sur-Yvette, France}
\affiliation{Max Planck Institute for the Physics of Complex Systems, N\"othnitzer Str. 38, 01187 Dresden, Germany}
\affiliation{SUPA, Physics Department, IPAM and Institute for Complex Systems and Mathematical Biology,
King's College, University of Aberdeen, Aberdeen AB24 3UE, United Kingdom}


\author{Anton Peshkov}
\affiliation{Service de Physique de l'Etat Condens\'e, CNRS URA 2464, CEA-Saclay, 91191 Gif-sur-Yvette, France}
\affiliation{Max Planck Institute for the Physics of Complex Systems, N\"othnitzer Str. 38, 01187 Dresden, Germany}
\affiliation{LPTMC, CNRS UMR 7600, Universit\'e Pierre et Marie Curie, 75252 Paris, France}

\author{Igor S. Aranson}
\affiliation{Materials Science Division, Argonne National Laboratory, 9700
  South Cass Avenue, Argonne, IL 60439}
\affiliation{Max Planck Institute for the Physics of Complex Systems, N\"othnitzer Str. 38, 01187 Dresden, Germany}

\author{Eric Bertin}
\affiliation{Laboratoire Interdisciplinaire de Physique, Universit\'e Joseph Fourier Grenoble, CNRS UMR 5588, BP 87, 38402 Saint-Martin d'H\`eres, France}
\affiliation{Universit\'e de Lyon, Laboratoire de Physique, ENS Lyon, CNRS, 46 all\'ee d'Italie, 69007 Lyon, France}
\affiliation{Max Planck Institute for the Physics of Complex Systems, N\"othnitzer Str. 38, 01187 Dresden, Germany}

\author{Francesco Ginelli}
\affiliation{SUPA, Physics Department, IPAM and Institute for Complex Systems and Mathematical Biology,
King's College, University of Aberdeen, Aberdeen AB24 3UE, United Kingdom}

\author{Hugues Chat\'{e}}
\affiliation{Service de Physique de l'Etat Condens\'e, CNRS URA 2464, CEA-Saclay, 91191 Gif-sur-Yvette, France}
\affiliation{Max Planck Institute for the Physics of Complex Systems, N\"othnitzer Str. 38, 01187 Dresden, Germany}
\affiliation{LPTMC, CNRS UMR 7600, Universit\'e Pierre et Marie Curie, 75252 Paris, France}

\date{\today}

\begin{abstract}
We show that ``dry" active nematics, {\it e.g.} collections of shaken elongated granular particles, 
exhibit large-scale spatiotemporal chaos
made of interacting dense, ordered, band-like structures 
in a parameter region including the linear onset of nematic order. 
These results are obtained from
the study of the relatively simple and well-known (deterministic) hydrodynamic equations describing these systems
in a dilute limit, and of a self-propelled particle Vicsek-like model for this class of active matter. In this last case, revisiting the status of the strong fluctuations and long-range correlations now considered as landmarks of orientationally-ordered active phases, we show that the “giant number fluctuations” observed in the chaotic phase are a trivial consequence of density segregation.  However anomalous density  fluctuations are present in the homogeneous quasi-ordered nematic phase and characterized by a non-trivial scaling exponent.
\end{abstract}

\maketitle

Many of the recent studies on active suspensions and active gels have reported the
existence of instabilities leading to spontaneous flows \cite{SR-REVIEW,GELS,CATES2,RMP}.
At the nonlinear level, the term ``bacterial turbulence'' has been used
to describe the fast, collective but chaotic motion of swimmers evolving
in what remains a very low Reynolds number, inertialess, world 
\cite{sa2012,GOLDSTEIN,Sokolov, SAINTILLAN}.
In remarkable in vitro experiments on actomyosin motility assays, the fluid
in which filaments and motors evolve seems to play a key role in the emergence of
local order leading in turn to erratic large-scale flows \cite{BAUSCH}.
Other actomyosin systems,
like the active nematics suspensions studied by Dogic {\it et al.}, display spontaneous
large-scale dynamics mediated by the nucleation and motion of topological defects
\cite{DOGIC,MCM-DEFECT,YEOMANS, CATES}.

In contrast, most of the recent studies of ``dry'' active matter (where the fluid in which
active particles move can be safely neglected), have not reported widespread occurence
of large-scale chaos or turbulence \cite{HOWEVER}.
Following the seminal papers of Vicsek {\it et al.} \cite{VICSEK} and Toner
and Tu \cite{TT}, a lot of attention has been paid to the nature of the onset of orientational
order/collective motion and to the existence of generic long-range correlations and anomalous
fluctuations in spatially-homogeneous ordered phases \cite{TTU,CHATE,TAILLEUR}.
The existence of
long-wavelength instabilities of homogeneous ordered states leading, in dilute systems,
to some phase separation between high-density high-order structures (bands, waves)
is now recognized as a generic feature \cite{CHATE, IGOR-LEV, falko,MCM, MCM2,
  bertin, RODS, Peshkov2012a, Peshkov2012b, MCM-polar, MCM2012},
but the  stability and large-scale dynamics of these structures remain largely unknown.

In this Letter, we show that large-scale spatiotemporal chaos arises
generically in dry active nematics. 
We first demonstrate that the solutions of the relatively simple and well-known (deterministic) 
hydrodynamic equations describing these systems
are chaotic in a region of parameter space including the linear onset of nematic order.
We show in particular that the nonlinear ordered band solution found
before \cite{NEMAMESO} is unstable, leading to
a disordered phase in which elongated dense and ordered structures curve, extend, merge, and split on 
very large time- and length-scales.
Returning to the Vicsek-style model for dry active nematics introduced in \cite{CGM}
to investigate further the effect of fluctuations,
we provide evidence that its segregated regimes where dense, ordered structures form
are also actually chaotic on large scales, rendering this phase asymptotically
disordered.
The giant number fluctuations reported in \cite{CGM} to be in agreement with the predictions
of Ramaswamy {\it et al.} \cite{GNF} are thus a trivial consequence of phase separation.
Non-trivial fluctuations are nevertheless present in the homogeneous quasi-ordered phase overlooked in
\cite{CGM} but found here at larger densities. Their scaling exponent, though, differs
from the simple one derived in \cite{GNF} using a
linearized theory. Interestingly, we find that it takes a value similar
to that calculated by Toner and Tu 
for {\it polar} ordered phases \cite{TT}.

We start by recalling the Vicsek-style model for active nematics defined in \cite{CGM},
where point particles carrying a (uniaxial) nematic degree of freedom
align locally and are forced to move randomly along one of the two
directions defined by their axis.
In two spatial dimensions the positions ${\bf x}_j^t$ and directors
${\bf n}^t_j \equiv \left(\cos \theta^t_j,\sin \theta^t_j\right)^T$ with
$\theta^t_j \in [-\frac{\pi}{2}, \frac{\pi}{2}]$ of
particles $j\!=\!1,\ldots,N$ are updated synchronously at discrete
timesteps according to:
\begin{equation}
\theta^{t+1}_j = \frac{1}{2} \mathrm{Arg}\left[ \sum_{k \in V_j} e^{i 2
\theta^t_k} \right] + \psi^t_j
 \; ;\;\;
{\bf x}_j^{t+1} =  {\bf x}_j^t \pm v_0 \,\hat{\bf n}_j^t
\label{eq:VM}
\end{equation}
where $V_j$ is the set of neighbors of particle $j$ within unit distance,
the sign in the second equation is chosen randomly with equal probability, and
the random angle $\psi^t_j\in[-\eta \frac{\pi}{2},\eta\frac{\pi}{2}]$ (with $\eta\in[0,1]$) is drawn from a uniform distribution.
As a matter of fact, this model has not been much studied beyond the
initial paper \cite{CGM} where numerical simulations performed 
on square domains of linear size $L$ at global density
$\rho_0\!=\!N/L²\!=\!\frac{1}{2}$ concluded to an isotropic/nematic Berezinskii-Kosterlitz-Thouless-like transition \cite{BKT}
as $\eta$ is decreased, with the quasi-long-range ordered phase consisting of
a single dense ordered band and supporting giant number fluctuations.

As shown in \cite{NEMAMESO},
the rather well known hydrodynamic equations for dry active nematics \cite{Ramaswamy} can be derived
in a simple and controlled way from this model, with all transport coefficients
depending explicitly on $\rho_0$ and the noise strength,
the only two parameters remaining after rescaling. In this approach,
one assumes a dilute limit and a molecular chaos hypothesis,
which allows to write a Boltzmann equation for the one-body distribution function
$f({\bf x},\theta,t)$. 
By expanding $f$ in Fourier series of $\theta$,
$f({\rm x},\theta,t) \!=\! \frac{1}{\pi} \sum_{k=-\infty}^{k=\infty} \hat{f}_k({\rm x}, t) e^{-i 2 k \theta}$,
the kinetic equation becomes a hierarchy which can be truncated and closed, assuming
a diffusive scaling ansatz and the proximity of the onset of nematic order. 
The first non-trivial order yields a nonlinear equation governing the nematic complex field
$Q \!\equiv\!\hat{f}_1$,
together with the continuity equation governing the density field $\rho\!\equiv\! \hat{f}_0$
\footnote{Eqs.(\ref{eq:RhoComplex}) and (\ref{eq:last}) can be written in terms of the more familiar
symmetric traceless tensor field ${\bf Q}$ since
$\rho [{\bf Q}]_{xx}=-\rho [{\bf Q}]_{yy}=\frac{1}{2}{\rm Re}\hat{f}_1$
and $\rho [{\bf Q}]_{xy}=\rho [{\bf Q}]_{yx}=\frac{1}{2}{\rm Im}\hat{f}_1$
but the complex notations are very convenient and we keep them in the following.}:
\begin{eqnarray}
\partial_{t}\rho &=&  \frac{1}{2}\Delta\rho+\frac{1}{2}{\rm Re}\left({\nabla^*}^{2}Q\right)
\label{eq:RhoComplex}\\
\partial_{t}Q &=& \left(\mu(\rho)-\xi\left|Q\right|^{2}\right)Q+\frac{1}{4}\nabla^{2}\rho+\frac{1}{2}\Delta Q
\label{eq:last}
\end{eqnarray}
where $\mu(\rho)\!=\!\mu^\prime(\rho-\rho_{\rm t})$ and we have used the complex operators
$\nabla\!\equiv\!\partial_{x}+i\partial_{y}$,
$\nabla^*\!\equiv\!\partial_{x}-i\partial_{y}$,
and $\Delta\equiv\nabla\nabla^*$.
The transport coefficients $\mu'$, $\rho_{\rm t}$, and $\xi$ are positive
constants depending on the noise strength $\sigma$
\footnote{They read
$\mu(\rho) \!=\!  \frac{8}{3\pi}\left[\left(2\sqrt{2}-1\right)\hat{P}_{1}-\frac{7}{5}\right]\rho-\left(1-\hat{P}_{1}\right)$ and
$\xi =  \frac{32 \nu}{35\pi^{2}}\left[\frac{1}{15}+\hat{P}_{2}\right]
\left[\left(1+6\sqrt{2}\right)\hat{P}_{1}-\frac{13}{9}\right]$
with $\nu \!=\!  \left[\frac{8}{3\pi}\left(\frac{31}{21}+\frac{\hat{P}_{2}}{5}\right)\rho_0+\left(1-\hat{P}_{2}\right)\right]^{-1}$, where the $\hat{P}_k$ are the coefficients of
the Fourier series of the noise distribution. In the following, we used a
Gaussian distribution of variance $\sigma^2$ so that $\hat{P}_k\!=\!e^{-2k^2\sigma^2}$.}.

\begin{figure}[t!]
\includegraphics[width=\columnwidth]{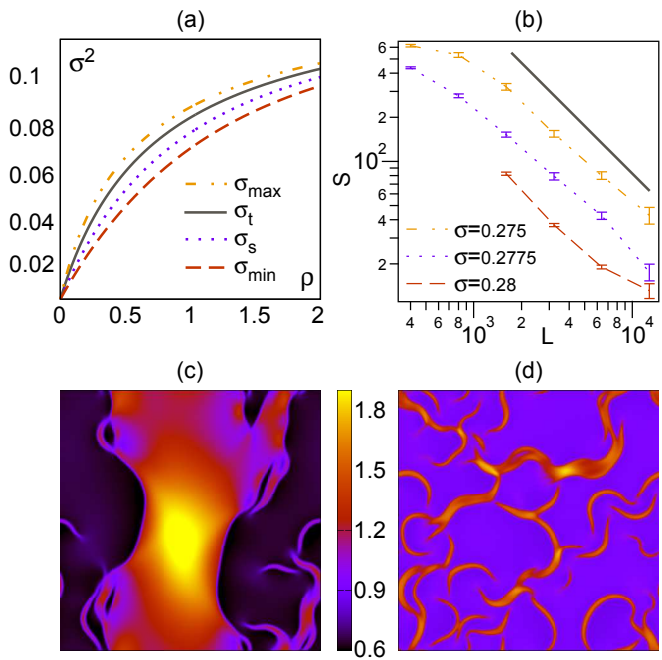}
\caption{(color online) Mesoscopic dynamics:
(a) phase diagram of hydrodynamic equations for active nematics.
(b) global order parameter vs system size at different noise values in chaotic regime  (the solid line has slope $-1$).
(c,d) snapshots of density field in chaotic regime for $L=1600$ and $\rho_0=1$; (c): localized chaos ($\sigma=0.26$); 
(d) fully-developed chaos ($\sigma=0.28$)
}
\label{fig1}
\end{figure}

The phase diagram of Eqs.(\ref{eq:RhoComplex}) and (\ref{eq:last}) is given in Fig.~\ref{fig1}a.
The condition $\rho\!=\!\rho_{\rm t}$, defining the line $\sigma_{\rm t}$,
marks the linear instability of the disordered solution $Q\!=\!0$
and the emergence of the homogeneous ordered solution $|Q|\!=\!\sqrt{\mu/\xi}$ (for $\mu\!>\!0$).
But this ordered solution is itself linearly unstable to long-wavelength perturbations
transversal to nematic order in a region bordering the basic line
$\rho_{\rm t},\sigma_{\rm t}$.
Deeper in the ordered phase, below the line $\rho_{\rm s},\sigma_{\rm s}$,
the homogeneous ordered solution is linearly stable.

It was shown in \cite{NEMAMESO} that Eqs.(\ref{eq:RhoComplex})
and (\ref{eq:last}) support an inhomogeneous solution in the form of a band of
nematic order with density $\rho_{\rm band}\!>\!\rho_{\rm s}$ surrounded by a disordered
gas with $\rho_{\rm gas}\!<\!\rho_{\rm t}$.
Supposing the nematic order is along $x$, for this  solution we obtain $\rho=R_0(y)\equiv Q_0(y)+ \rho_{\rm gas}$ and 
\begin{eqnarray}
\label{eq:f1-band}
Q_0(y)&=&  \frac{ 3 (\rho_t - \rho_{\rm gas})}
{1+a\cosh\left(\sqrt{4\mu^\prime (\rho_t-\rho_{\rm gas})} y\right)}
\end{eqnarray}
with $a\!=\!\sqrt {1-9 \xi (\rho_t -\rho_{\rm gas}) /2\mu^\prime  }$, 
$\rho_{\rm gas}$ being a constant fixed by density conservation \cite{NEMAMESO}.
Its existence domain
($\sigma_{\rm min},\sigma_{\rm max}$) actually extends beyond the
region of linear instability of the homogeneous ordered solution
($\sigma_{\rm min}\!<\!\sigma_{\rm s}$ and $\sigma_{\rm max}\!>\!\sigma_{\rm t}$).
Its ordered part occupies a {\it fraction} of the $y$ dimension of the system going
continuously from zero (near $\sigma_{\rm max}$) to one (near $\sigma_{\rm min}$).

If considered only as a one-dimensional function of $y$, the band solution
 Eq.~(\ref{eq:f1-band}) is linearly stable.
We now show that it is always unstable
with respect to long-wavelength undulations along the $x$-axis.
To study the linear stability with respect to
wavenumber $k$, we seek the perturbative solution in the form $(  Q, \rho) \!=\! (Q_0(y), R_0(y)) + (q(y),r(y)) \exp(\lambda t + i k x)$.
Substituting the growth rate $\lambda$ into Eqs. (\ref{eq:RhoComplex}),(\ref{eq:last}),
we obtain a linear system for $q(y), r(y)$.
Noting that for $k=0$ the solution to this system is the translational mode
$r\!=\!q\! =\! \partial_y Q_0$, we can further simplify the problem in the long-wave limit $k\to 0$
by expanding the perturbative solution in $k$ and employing the following anzatz:
$\lambda\!=\! \lambda_1 k^2$, $q(y)=u(y)+ i v(y) $  and
\begin{eqnarray}
\left(
\begin{array}{c}
 r(y)  \\
 u(y)   \\
  v(y)
\end{array}
\right)
= \left(
\begin{array}{c}
\partial_y Q_0(y)   \\
 \partial_y Q_0(y) \\
0
\end{array}
\right)
+
\left(
\begin{array}{c}
k^2  r_1 (y)   \\
k^2  u_1 (y) \\
i k v_1 (y)
\end{array}
\right)
\label{expansion}
\end{eqnarray}
We then obtain linear inhomogeneous self-adjoint equations for the
functions $r_1$, $u_1$, and $v_1$.
The solvability condition yields an explicit expression
for the growth rate $\lambda$ in terms of integrals of $Q_0$ and $\partial_y Q_0$.
The analysis, detailed in \cite{EPAPS},
shows that $\lambda\!>\!0$, implying that the band solution is always unstable.
However, the instability can be suppressed in small systems.

At the nonlinear level, for large enough systems,
the instability of the band first manifests itself as some
periodic modulation in space and time localized along its borders. This then turns into
localized chaotic behavior (Fig.~\ref{fig1}b), which eventually
develops into full-blown spatiotemporal chaos for large enough system sizes and integration times.
There, distorted band-like structures evolve
on very long timescales and large lengthscales,
elongate, split, merge, without ever forming the original macroscopic band again
(Fig.~\ref{fig1}c and \cite{EPAPS}).
We have observed this spectacular dynamics all along the
($\sigma_{\rm min},\sigma_{\rm max}$) interval \footnote{This was done at a global density
$\rho_0=1$, but we do not have any reason to believe that a qualitative change
of behavior occurs as $\rho_0$ is varied.}.
As $\sigma$ is varied from $\sigma_{\rm max}$ to $\sigma_{\rm min}$, the largest
structures observed have increasing sizes. 
We measured the global nematic order parameter
$S(t)\!=\!|\langle Q \rangle_{x,y}|$ and the
two-point spatial correlation function of the density field for different
square systems of linear size $L$.
For large enough systems, the time-averaged order parameter $\langle S\rangle$ decreases like $1/L$,
indicating the existence of a finite, $L$-independent correlation length (Fig.~\ref{fig1}d)
\footnote{This is corroborated by the two-point correlation function of the density field: it is axisymmetric for long enough averaging times and 
its radial average decreases exponentially, with a system-size independent cutoff for large enough $L$ 
(not shown).}.
The segregated phase of the hydrodynamic equations for active nematics is thus
asymptotically disordered.

We now investigate the robustness of the above results with respect to fluctuations
by coming back to the Vicsek-like model defined by Eq.~(\ref{eq:VM}). This also provides
an opportunity to gauge the faithfulness
---at a qualitative level---
of the hydrodynamic equations
(\ref{eq:RhoComplex})-(\ref{eq:last}) to the model they were derived from.
We performed extensive simulations of Eq.~(\ref{eq:VM}) at various global densities $\rho_0$,
varying the noise strength $\eta$ and the system size $L$.

\begin{figure}[t!]
\includegraphics[height=4cm]{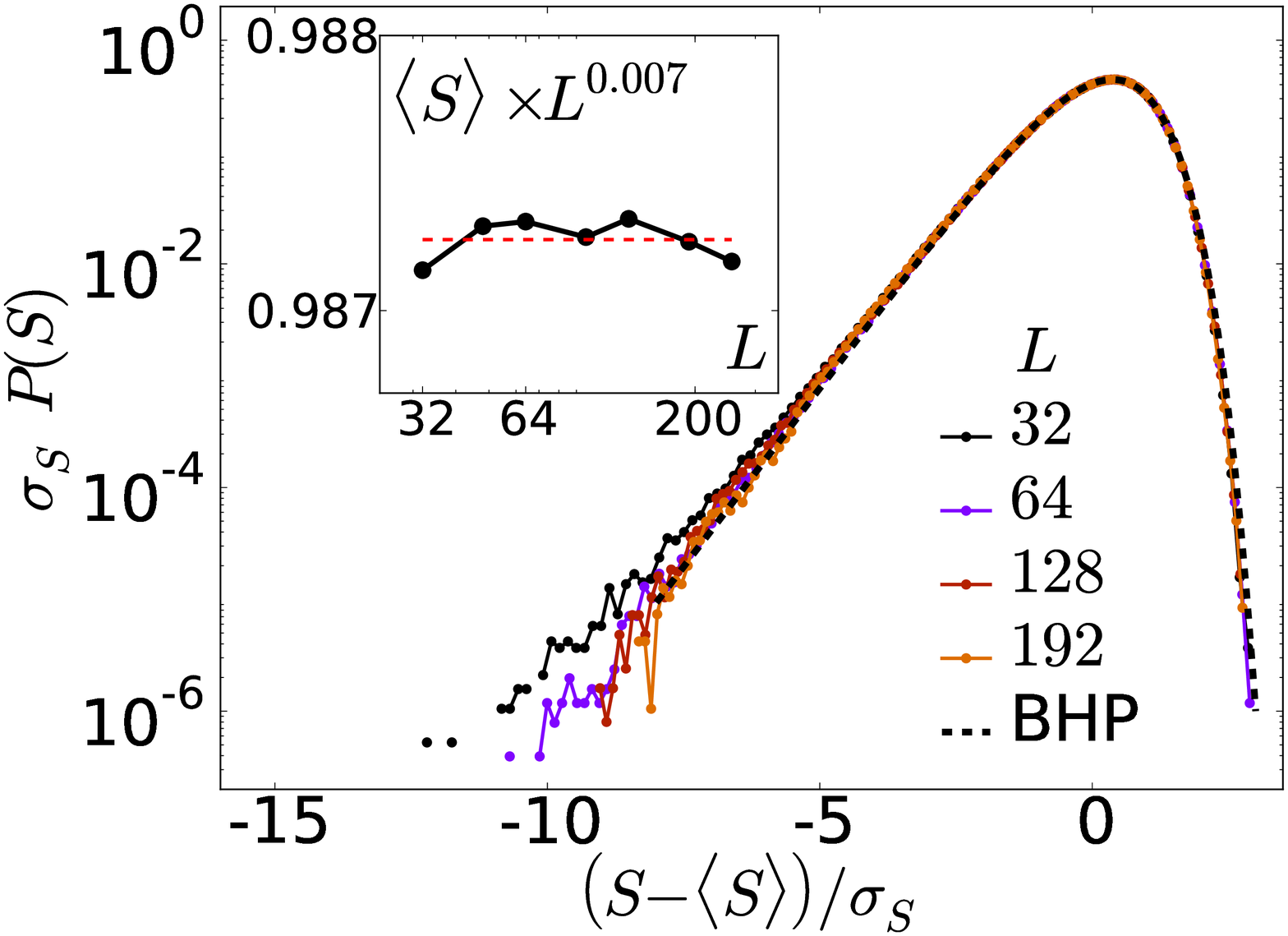}
\includegraphics[height=4.cm]{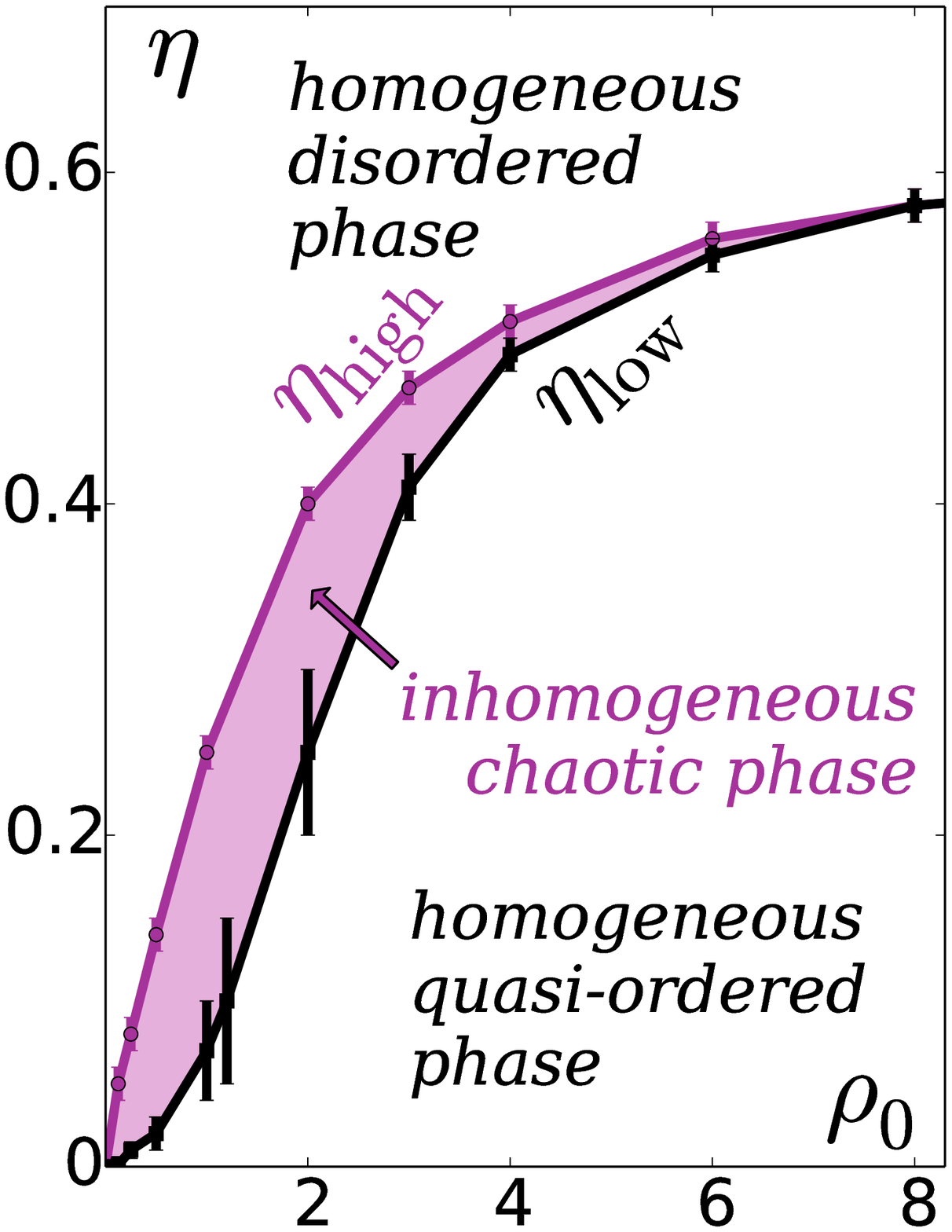}\vspace{0.4cm}
\includegraphics[height=4cm]{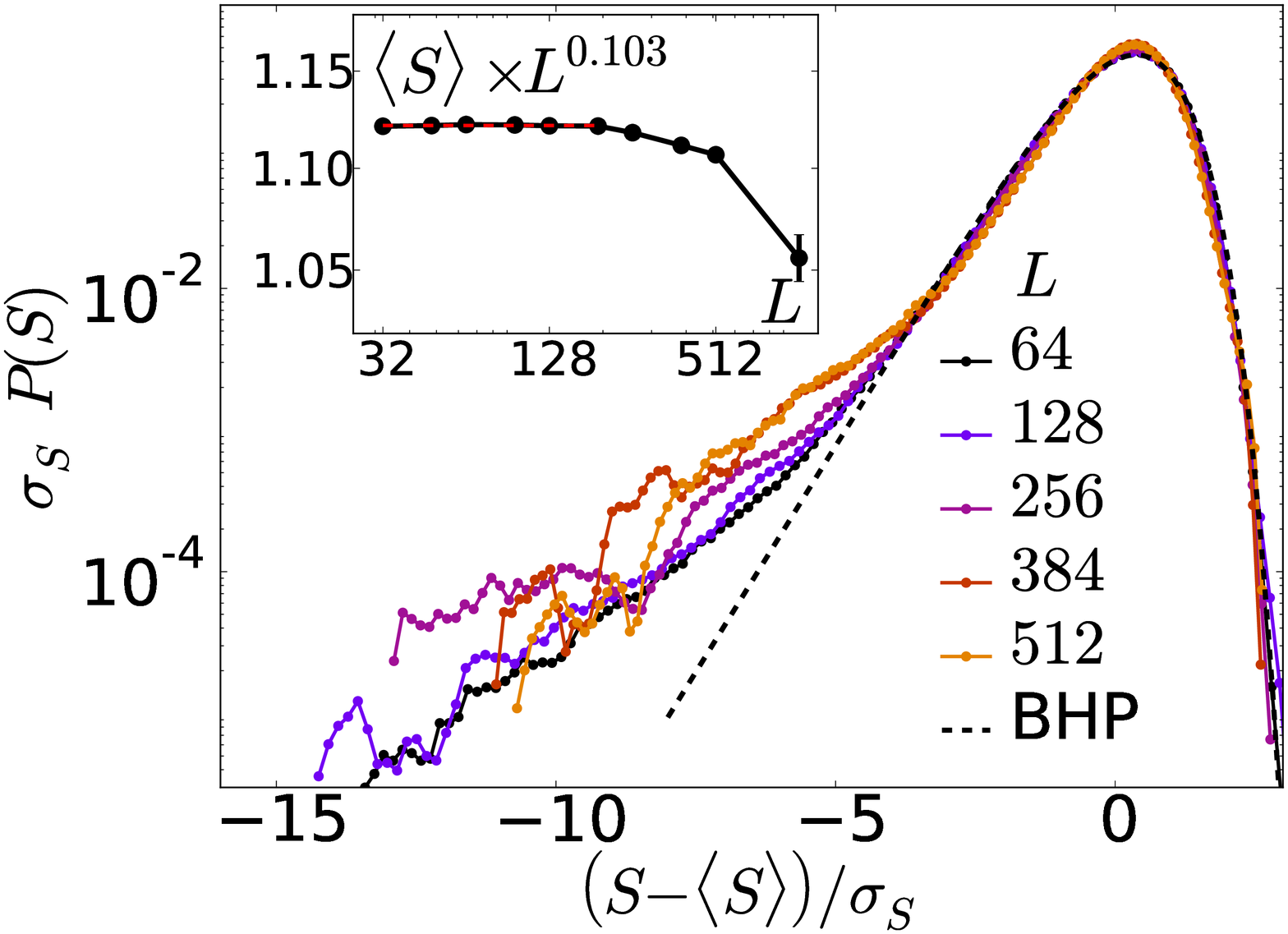}
\includegraphics[height=4.cm]{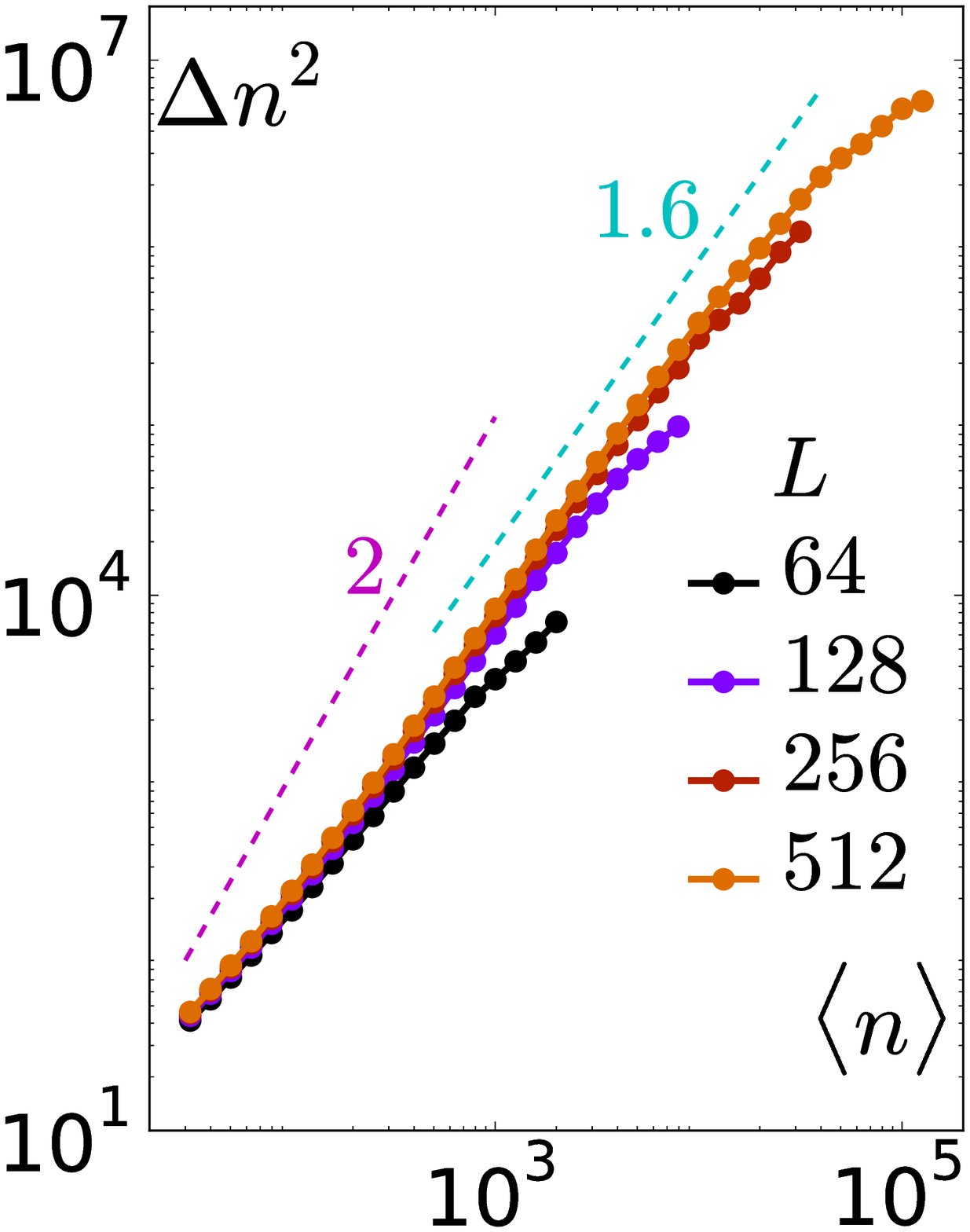}\vspace{0.3cm}
\hspace*{0.45cm}\includegraphics[width=4.cm]{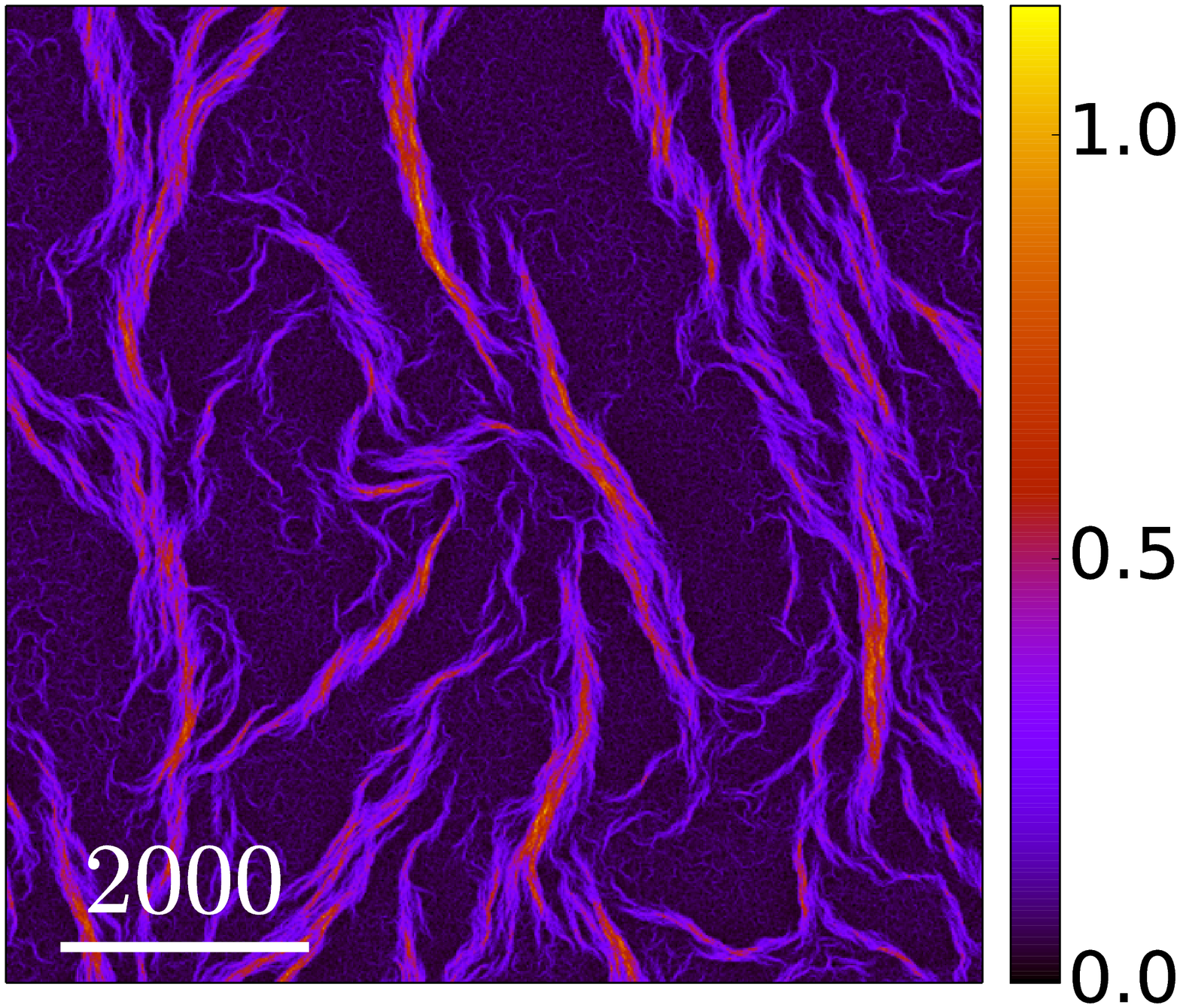}
\includegraphics[width=4.cm]{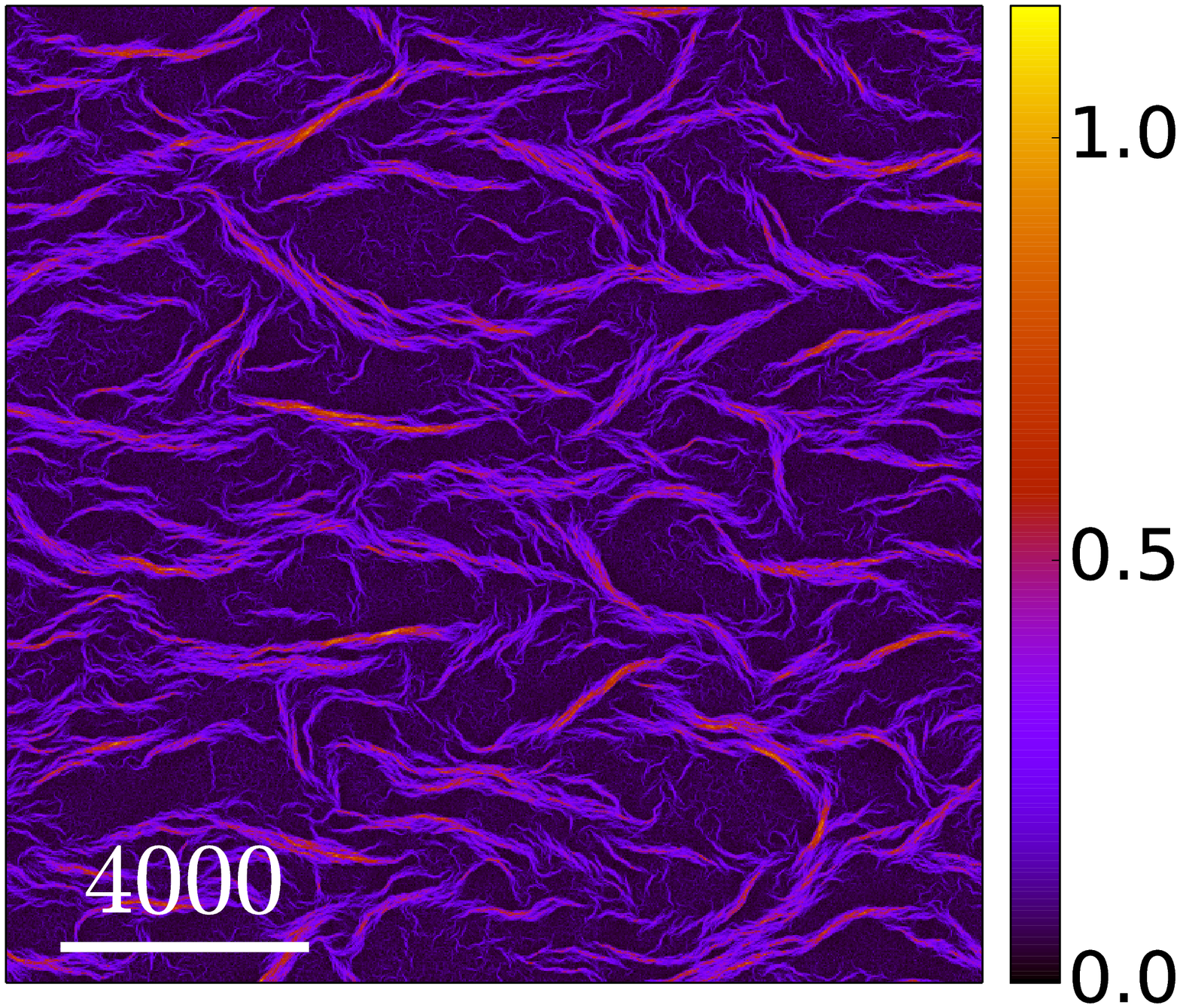}
\begin{picture}(0,0)
\put(-150,356){(a)}
\put(-40,356){(b)}
\put(-150,228){(c)}
\put(-40,228){(d)}
\put(-180,103){(e)}
\put(-70,103){(f)}
\end{picture}
\caption{(color online)
Microscopic dynamics: (a,c) Rescaled distribution of $S$ at various system sizes in the homogeneous ordered (a) and inhomogeneous (c) phases.
Insets of (a,c): log log plots of $\langle S \rangle \times L^{-\zeta}$ vs $L$, where $\zeta$ is estimated
from a fit of the initial decay of  $\langle S \rangle$. 
(b) Phase diagram in the $(\eta,\rho_0)$ plane.
(d) Variance $\Delta n^2 $ as a function of mean number of particle
 $\langle n \rangle$, in the homogeneous ordered phase. Dashed lines: 
 power laws with exponents $2$ (magenta) and $1.6$ (cyan).
(e-f) Snapshots of coarse-grained density in the chaotic phase of
panel (c) for system sizes $L\!=\!8192$ (e) and $L\!=\!16384$ (f).
Parameters: in all cases $v_0=0.3$,  for (a,d): $\rho_0\!=\!2$, $\eta\!=\!0.1$, $\zeta\!=\!0.007$ 
and for (c,e,f): $\rho_0\!=\!\frac{1}{8}$, $\eta\!=\!0.038$, $\zeta\!=\!0.103$.
}
\label{fig2}
\end{figure}

At large enough $\rho_0$, we do observe, for low enough $\eta$,
a spatially-homogeneous, non-segregated, quasi-ordered phase (not shown).
Because it possesses
unusually strong density fluctuations (see below) which may be hard to distinguish from
the fluctuating structures of the segregated phase, the location of $\eta_{\rm low}$,
the noise value marking the lower extent of the inhomogeneous phase,
is difficult to characterize beyond visual inspection provided
by movies and snapshots such as in Fig.~\ref{fig2} and \cite{EPAPS}. We used scaling and fluctuation
properties of the global nematic order parameter $S$, 
whose full probability distribution $P(S)$
can be measured with
good statistics only for moderate system sizes (up to $L\!=\!256$).
In the homogeneous phase, its mean $\langle S\rangle$ decreases algebraically with $L$ with an
exponent $\zeta(\eta)\!<\!\frac{1}{8}$ (quasi-long-range order),
and $P(S)$
quickly converges,
as $L$ is increased, to the Bramwell-Holdsworth-Pinton (BHP) distribution, well-known
to describe almost perfectly the quasi-ordered, vortex-free phase of the equilibrium XY model
\cite{XY} (Fig.~\ref{fig2}a). The inhomogeneous segregated phase, by contrast, is characterized
by a departure of $P(S)$ from the BHP distribution which becomes more important as the system
size is increased (Fig.~\ref{fig2}c). During the corresponding events,
the main dense ordered band typically observed at such moderate system sizes
reorganizes itself.
We used the finite-size behavior of $P(S)$
to define ---admittedly rather roughly--- the threshold value $\eta_{\rm low}$: for
 $\eta\!<\!\eta_{\rm low}$,  $P(S)$ falls on the BHP distribution, whereas
for $\eta\!>\!\eta_{\rm low}$  $P(S)$ deviates from BHP, and
these deviations eventually become so important that the initial algebraic decay of $\langle S\rangle$
with $L$ accelerates at large $L$ values (inset of Fig.~\ref{fig2}c).

Using the above approach, we find that $\eta_{\rm low}$,
decreases with $\rho_0$ (Fig.~\ref{fig2}b).
For $\rho_0\!=\!\frac{1}{2}$, the density used in \cite{CGM}, we estimate
$\eta_{\rm low}\!\simeq\! 0.02$, relegating the homogeneous regime to numerically difficult,
very small noise strengths regimes not probed in \cite{CGM}.
For even smaller $\rho_0$ values the homogeneous phase is practically unobservable.

For the $\eta_{\rm high}$ line of the phase diagram
separating the segregated from the homogeneous
{\it disordered} phases, we used the location of the peak of the susceptibility of the nematic order
parameter. Even though both phases are disordered, the segregated one still shows sizeable $S$ values
at the finite sizes we can probe numerically.

Our numerical results obtained for $\eta_{\rm low}\!<\!\eta\!<\!\eta_{\rm high}$ show that
the decay of $\langle S\rangle$ accelerates at large system sizes as $P(S)$ gradually departs from the BHP distribution.
We thus conjecture that our system is then, asymptotically, in a {\it disordered}, not a quasi-ordered phase.
This occurs even when the initial decay of $\langle S\rangle$ displays an exponent $\zeta\!<\!\frac{1}{8}$,
its value at the Berezinskii-Kosterlitz-Thouless transition
used in \cite{CGM} to define the isotropic/nematic transition  (inset of Fig.~\ref{fig2}c).
In particular, at the onset of density segregation ($\eta\!\simeq\!\eta_{\rm low}$),
the decay exponent $\zeta$ is typically rather small
(e.g. $\zeta \!\simeq \! 0.03$ for $\rho_0\!=\!0.5$).

Observing full-blown disordered inhomogeneous regimes for $\eta$ values such that the
early decay of $\langle S\rangle$ occurs with an exponent $\zeta\!<\!\frac{1}{8}$ is however
numerically very difficult, requiring huge system sizes and simulation times.
The snapshots shown in Fig.~\ref{fig2}e-f are actually taken in such
a case: several dense, curved bands of various orientations are present.
They evolve on very long timescales, elongating, merging, splitting, in a manner
reminiscent of the band chaos reported in Fig.~\ref{fig1} (see \cite{EPAPS}). 
Only the largest system sizes (Fig.~\ref{fig2}f) reveal
the existence of a finite characteristic length, in agreement with the nonlinear analysis of
the equations (\ref{eq:RhoComplex})-(\ref{eq:last}).

Next, we investigated anomalous, ``giant'' number fluctuations, which we characterize by
the scaling of $\Delta n^2$,
the variance of the number $n$ of particles contained in a square sub-system. Even though this
anomalous scaling is a landmark of fluctuating active ordered phases, it can be probed in the
fluctuation-free context of the fully chaotic regime of the hydrodynamic equations
(\ref{eq:RhoComplex}) and (\ref{eq:last}).
We find $\Delta n^2\!\sim\! n^2$ for boxes of sizes up to the characteristic lengthscale of chaos,
followed by a crossover to normal fluctuations ($\Delta n^2 \! \sim \!n$) for larger boxes
(not shown). For the inhomogeneous disordered phase
of the Vicsek-like model (\ref{eq:VM}), as already reported in \cite{CGM},
we also find $\Delta n^2 \!\sim\! n^2$ but are numerically unable to see the expected crossover
to normal fluctuations. These observations are the trivial result of the segregation of
density in dense bands \cite{Aranson} and not due to the sophisticated mechanism put forward by Toner,
Ramaswamy, {\it et al.}
as a typical feature of ordered, non-segregated, active phases.
However, the homogeneous regimes found for $\eta\!<\!\eta_{\rm low}$ in the microscopic model
do constitute such a phase. Measuring number fluctuations in this case,
we find anomalous fluctuations, $\Delta n^2\!\sim \! n^\alpha$,
but with a characteristic scaling exponent $\alpha$ close to 1.6, not 2
(Fig.~\ref{fig2}d). In other words,
we do {\it not} find the value $\alpha\!=\!2$ derived by Ramaswamy {\it et al.} from a linearized
theory, but, surprisingly,
a value close to that calculated by Toner and Tu for active {\it polar} ordered phases.

We finally mention results obtained on the Vicsek-like model defined by Eq.~(\ref{eq:VM}), but where
neighbors are chosen to be those forming the first shell of Voronoi
polygons around a given particle. 
In this ``metric-free" model, the basic instability of the homogeneous ordered state leading to 
the segregated phase is suppressed \cite{Peshkov2012b}. Accordingly, our simulations 
reveal only two phases, the homogeneous disordered one, and the
homogeneous quasi-ordered one, separated by a Berezinskii-Kosterlitz-Thouless-like
transition. 
The quasi-ordered phase also exhibits anomalous number fluctuations,
also with a non-trivial scaling exponent $\alpha\!\simeq\!1.6$ (not shown). 

To summarize, we showed that dry active nematics exhibit large-scale spatiotemporal chaos
consisting of moving, elongating, splitting, merging high-density high-order band-like structures.
At the level of the (deterministic) hydrodynamic equations (\ref{eq:RhoComplex}) and (\ref{eq:last}),
this chaos is the outcome of the linear instability, in two dimensions, of the band solution
(\ref{eq:f1-band}), and is observed in the whole region of existence of the solution,
{\it i.e.} between the  $\sigma_{\rm min}$ and $\sigma_{\rm max}$ lines
in Fig.~\ref{fig1}a, a region which encompasses the linear onset of nematic order.
At the level of the Vicsek-like microscopic model defined by (\ref{eq:VM}),
the corresponding lines are the $\eta_{\rm low}$ and
 $\eta_{\rm high}$ lines of  Fig.~\ref{fig2}b.
The phase diagram of dry active nematics
thus comprises three different phases, a homogeneous disordered phase at strong noise, an
inhomogeneous disordered chaotic phase at intermediate noise values, and
a homogeneous ordered phase.
In the microscopic model, this last phase is
only quasi-ordered and displays anomalous number fluctuations.
The order-disorder transition is thus located between the ordered phase and the chaotic inhomogeneous
phase. Its nature, even at the ``mean-field'' level of the hydrodynamic equations, remains unclear:
if the correlation scales of chaos were found to diverge when $\sigma\to\sigma_{\rm min}$,
then the transition would probably be continuous.
Unfortunately, the numerical determination of these scales in this limit
is a very difficult task beyond the scope of this Letter
\footnote{Note that this would be an even harder task 
with the microscopic model.}.

Our results should eventually be completed by a study of the mesoscopic theory derived in
\cite{NEMAMESO}, either by numerical integration of these Langevin equations or by some
renormalization group analysis.
(This could in particular provide some explanation to the anomalous scaling exponent
for number fluctuations found here similar to that of polar ordered phases.)
This difficult task is left for future studies. 
At the experimental level, we believe that microtubule motility assays of the type studied in 
\cite{SUMINO} should be able to demonstrate the phenomena described here \cite{Private}.

\acknowledgments
We thank the Max Planck Institute for the Physics of Complex Systems, Dresden,
for providing the framework of the Advanced Study Group
``Statistical Physics of Collective Motion''
within which much of this work was conducted. FG and SN acknowledge support from grant
EPSRC First Grant EP/K018450/1. The work of ISA was supported by the US Department of Energy, Office of Basic Energy Sciences, Division of Materials Science and Engineering, under contract no. DEAC02-06CH11357.

\end{document}